\pdfoutput=1
\documentclass[prl, twocolumn, superscriptaddress, nofootnoteinbib, showpacs,amsmath,amssymb]{revtex4}

\usepackage{float}
\usepackage{graphicx}

\begin{document}

\title{Coupling of nitrogen-vacancy centers in diamond to a GaP waveguide}
\author{K.-M.C. Fu}
\email{kai-mei.fu@hp.com}
\author{C. Santori}
\author{P.E. Barclay}
\affiliation{Information and Quantum System Lab, Hewlett-Packard
Laboratories, 1501 Page Mill Road, MS1123, Palo Alto, California
94304, USA}
\author{I. Aharonovich}
\author{S. Prawer}
\affiliation{School of Physics, The University of Melbourne, Melbourne, Victoria 3010, Australia}
\author{N. Meyer}
\author{A. M. Holm}
\affiliation{Hewlett-Packard Company, 1000 NE Circle Blvd., Corvallis, Oregon 97330, USA}
\author{R.G. Beausoleil}
\affiliation{Information and Quantum System Lab, Hewlett-Packard
Laboratories, 1501 Page Mill Road, MS1123, Palo Alto, California
94304, USA}

\begin{abstract}

The optical coupling of guided modes in a GaP waveguide to nitrogen-vacancy (NV) centers in diamond is demonstrated.  The electric field penetration into diamond and the loss of the guided mode are measured.  The results indicate that the GaP-diamond system could be useful for realizing coupled microcavity-NV devices for quantum information processing in diamond.

\end{abstract}
\maketitle

In recent years much progress has been made toward the realization of quantum information processing (QIP) based upon nitrogen-vacancy (NV) centers in diamond.  Long 350~$\mathrm{\mu s}$ ground state electron spin coherence times have been observed~\cite{Gaebel06a}, full electron spin control has been achieved using optically-detected magnetic resonance~\cite{ref:jelezko2004obs}, and electron-nuclear qubit transfer, necessary for long quantum memory times, has been performed~\cite{GurudevDutt07a}.  However, such demonstrations so far have involved manipulation only of isolated NV centers.  For realization of large-scale QIP~\cite{Benjamin06a} or for quantum repeater systems~\cite{Childress05a} it will be necessary to connect NV centers together through ``flying'' qubits such as photons.  To achieve this, optical structures in diamond such as microcavities and waveguides are needed to enable transfer of quantum information between the electron spin of the NV center and a photon. Optical structures in diamond could also be useful in enhancing photon collection efficiencies in recently proposed diamond-based magnetometers~\cite{ref:Taylor2008hig}.

To date, two approaches have been pursued to couple NV centers to an optical device.  The first is to couple NV centers in a diamond nanoparticle to a silica microcavity~\cite{Park06a}.  Unfortunately, nanoparticle NV centers exhibit spectral diffusion and large inhomogeneous broadening~\cite{Shen08a} making any practical, scalable device difficult to engineer.  Groups have also worked on fabricating optical structures out of diamond~\cite{Olivero05a,Wang07a}.   Although this second approach is promising, the coupling of NV centers to a diamond optical device has not yet been demonstrated.  In this work, we fabricate optical waveguides from high-index GaP (n$_\mathrm{GaP}$=3.3) thin films on top of diamond (n$_\mathrm{D}$=2.4) and demonstrate the coupling of the guided light to NV centers in the top 100~nm of the diamond.  We characterize the coupling strength of the guided mode to the surface NV centers as well as the scattering loss of the waveguides and calculate the expected NV coupled GaP ring-cavity performance based on these results.

\begin{figure*}
{\includegraphics[height = 40 mm,
keepaspectratio]{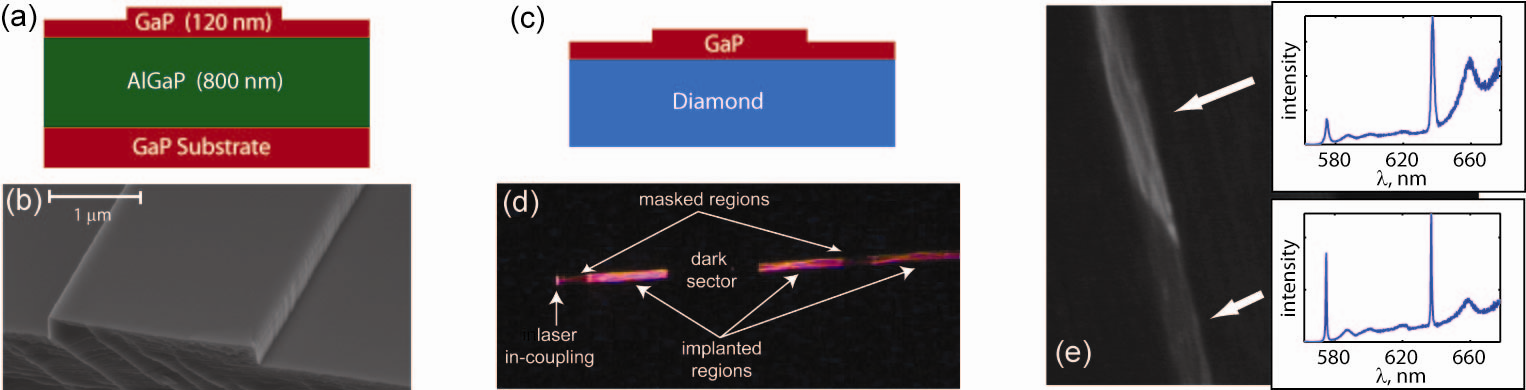}}\caption{(a) Waveguide structure before epitaxial liftoff. (b) SEM image of an etched 2~$\mathrm{\mu m}$-wide waveguide before liftoff.  ECR etch conditions: Ar/BCl$_3$/Cl$_2$ gases with 15/10/2 sccm flow, 200 W microwave power, 45 W RF power.   (c) Waveguide structure after liftoff and adhesion to diamond. (d) Optical excitation of NV centers through an 8~$\mathrm{\mu m}$ waveguide.  Implantation regions defined by the TEM grid and the dark sector boundary are clearly observed in the NV PL. (e) Sector boundary observed in the waveguide-excited PL in HPHT2.  Spectra taken with 532~nm excitation from the top of the waveguide show the 637~nm NV${^-}$ and 575~nm NV$^0$ ZPL lines in both sectors.}
\label{fig:images}
\end{figure*}

The initial GaP wafer consisted of a commercially available (IQE) MBE-grown 250~nm GaP epilayer on an 800~nm Al$_{0.8}$Ga$_{0.2}$P sacrificial layer on a GaP substrate.  The GaP epilayer was first thinned by ECR dry etching to 120~nm to increase the penetration depth of the evanescent field once the waveguides have been transferred to diamond.  Optical ridge waveguides ranging in width from 1-8 $\mathrm{\mu m}$ with a final ridge height of 50~nm were defined using standard optical lithographic and dry etching techniques (Fig.\ref{fig:images}a,b).  A 200~nm PECVD silicon nitride film was deposited on some of the waveguide samples.  The GaP epitaxial membrane was then undercut and removed from the GaP substrate and transferred to diamond in a procedure adapted from Ref.~\cite{Yablonovitch90a}. In this procedure, a 7:100 HF(49\%):water solution at room temperature was used to etch the Al$_{0.8}$Ga$_{0.2}$P sacrificial layer.  The measured etch rate was 51~$\mathrm{\mu m}$/hour along the $\langle100\rangle$ planes.

%
%
%
%

The diamond samples were Sumitomo high-pressure high-temperature grown diamond with a typical nitrogen content of 30-100~ppm.  The samples were implanted either with 2~MeV alpha particles with a dose of $10^{16}$~cm$^{-3}$ (HPHT1) , 200 keV Ga ions with a dose of $3\times10^{12}$~cm$^{-2}$  (HPHT2), or 30 keV Ga ions with a dose range of $10^{13}-10^{15}$~cm$^{-3}$ (HPHT3) in order to create vacancies. After implantation the samples were annealed at 925C in an H$_2$/Ar forming gas during which vacancies combined with already-present nitrogen impurities to create NV centers~\cite{Davies92a}.

The observation of NV photoluminescence (PL) from the implanted regions when exciting at 637~nm through the waveguide confirms the evanescent coupling of the guided electromagnetic field to NV centers near the surface.  In Fig.~\ref{fig:images}d 637~nm light from a continuous-wave tunable diode laser is end coupled through a high-NA lens into an 8 $\mathrm{\mu m}$-wide waveguide on HPHT1 at room temperature. The waveguide crosses a dark diamond sector as well as several implantation regions.  A long wavelength pass filter blocks scattered excitation light from the imaging CCD, and only light from the NV phonon sidebands with wavelength $\lambda>653$~nm is imaged.  At cryogenic temperatures, different sectors in HPHT diamond often exhibit different NV linewidths and peak intensities.  In Fig.~\ref{fig:images}e the waveguide coupling was performed at 10~K and a large contrast in intensity between the two diamond sectors is observed when the laser is slightly detuned from the narrower NV ZPL line in the lower sector.

\begin{figure*}
{\includegraphics[height = 40 mm,
keepaspectratio]{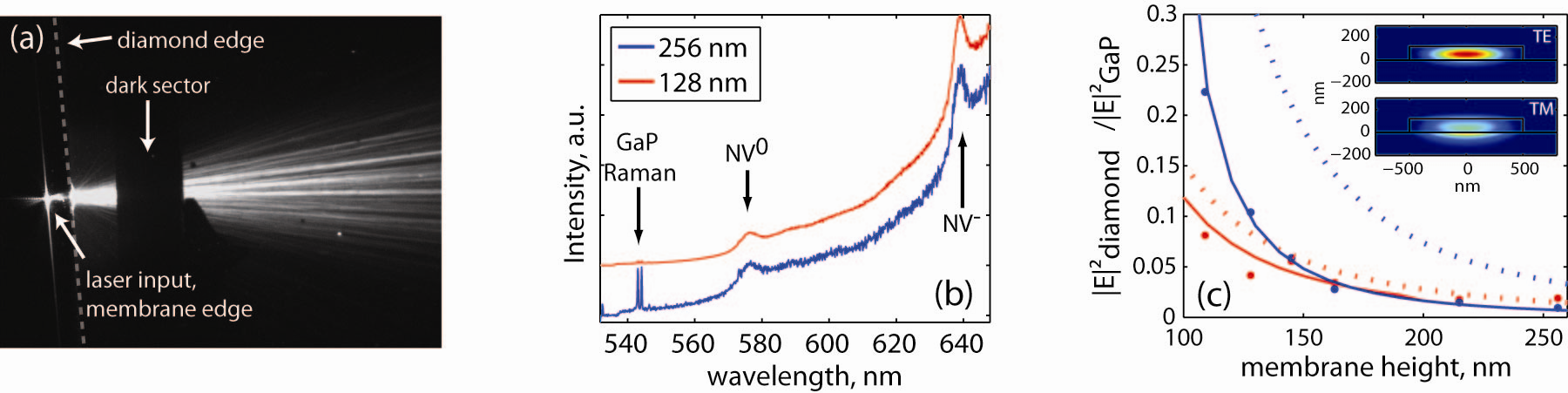}}\caption{(a) A TE-polarized 532 nm laser is coupled into the overhanging region of a 250~nm GaP membrane on diamond and NV PL is imaged on a CCD camera. (b) Representative spectra collected from the top of the membrane for membrane thicknesses of 256 and 128~nm for TM excitation. (c) The ratio of the electric field strength in the top 100 nm of the diamond and in the GaP waveguide for various membrane thicknesses. \emph{Red (Blue) dots}: TE (TM) experiment. \emph{Dotted lines}: \emph{Red (Blue)} TE (TM) calculation for a GaP waveguide on top of diamond. \emph{Solid lines}: \emph{Red (Blue)} TE (TM) calculation which includes a 4.7~nm air gap between the GaP and diamond.  \emph{Inset}: Simulated TE/TM electric field intensity in a 1~$\mathrm{\mu m}$-wide, 120~nm thick waveguide (no air gap).} \label{fig:thinning}
\end{figure*}

The two parameters critical to waveguide-NV coupled devices are the NV-waveguide coupling strength and the waveguide scattering loss.  The coupling strength depends on the per-photon electric field strength of the guided mode at the NV site, and good coupling requires that the gap between the GaP membrane and the diamond surface is minimized. In order to determine the coupling strength the following experiment was performed.  First a 250~nm GaP membrane was transferred to sample HPHT2. 532~nm laser light was end-coupled to the GaP membrane (Fig.~\ref{fig:thinning}a) and spectra containing both the NV PL and the GaP LO phonon Raman signal were collected from the top with a 10x objective for TE and TM excitation (Fig.~\ref{fig:thinning}b). To account for the relative brightness between the NV and GaP Raman signals, spectra were also obtained with top excitation from which a normalization factor was obtained that accounted for the GaP membrane thickness, the NV implantation depth (approximately 100~nm for 200~keV Ga ions), and the relative GaP Raman scattering efficiency between top and side excitation~\cite{ref:Loudon1964ram}.  The membrane was then thinned via an ECR dry etch and the optical measurements were repeated.  As seen in Fig.~\ref{fig:thinning}b, the ratio between the NV and the GaP Raman signals increased as the membrane was thinned and the evanescent field strength in the diamond increased.  For comparison, finite element simulations (COMSOL) were performed for a 1~$\mathrm{\mu m}$ wide waveguide to determine the relative electric field intensity in the waveguide and in the first 100~nm of the diamond for the lowest order TE and TM modes (Fig.~\ref{fig:thinning}c inset). The experimental and theoretical results for the dependence of the electric field strength ratio in the diamond and GaP as a function of membrane thickness are shown in Fig.~\ref{fig:thinning}c. Good agreement between the experimental data and the theoretical model can be obtained if a 4-5~nm air gap layer between the GaP and diamond is included in the model.  This small gap causes only a modest reduction in the electric field strength for the TE mode.  Currently we are unable to investigate the uniformity and composition of the diamond/GaP interface directly and can only say it is greater than the 1~nm diamond surface roughness measured by AFM on a representative Sumitomo HPHT sample.
%
%
%
%

The second parameter critical to GaP-NV diamond devices is the loss in the GaP at 637~nm.  We first measured this loss for waveguides on the original GaP wafer (similar to Fig.~\ref{fig:images}a). Fully etched waveguides with a waveguide height of 250~nm were fabricated with 6 lengths ranging from 0.1 to 3.2~mm.   A 637~nm laser was coupled into the end of each waveguide through a 0.95 NA objective and the scattered output was collected from the top using a 10$\times$ objective.  No dependence on waveguide width within the measurement uncertainty was observed indicating sidewall roughness is not the dominant source of loss.  The measured loss was $46\pm3$~dB/cm ($63\pm5$~dB/cm) for the TE (TM) mode.  This sample was next thinned to 160~nm and remeasured.  The TE mode loss was unchanged, however the TM loss increased to $80\pm6$~dB/cm.  The TM mode has a higher electric field intensity at the top and bottom waveguide surface than the TE mode which could account for the increased TM loss with the waveguide thinning.

%
%
\begin{figure}
{\includegraphics[height = 40 mm,
keepaspectratio]{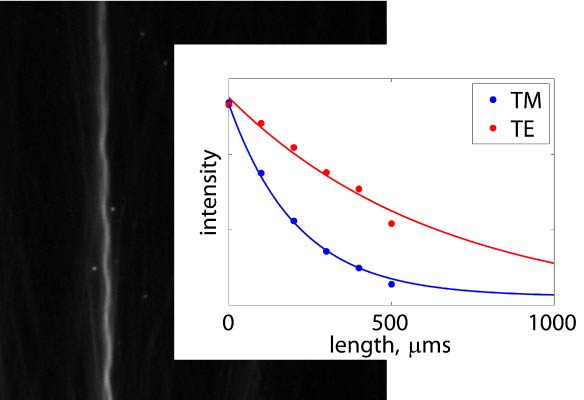}}\caption{(a) NV PL from a 1~micron waveguide excited at 637 nm on sample HPHT3.  (b) Decay of the PL in the 1 micron waveguide due to scattering loss.  The measurement length is limited to the length of the diamond sector.} \label{fig:loss}
\end{figure}
%
%

The loss for GaP waveguides once transferred to diamond was tested by exciting from the end of a ridged waveguide and measuring the intensity of the NV PL collected from the top of the waveguide in a single diamond sector.  The longest decay lengths were obtained for waveguides which included a silicon nitride cladding layer on top of the GaP.  A representative decay curve for a 1~$\mathrm{\mu m}$~wide waveguide is shown in Fig.~\ref{fig:loss}.  Loss for the TE (TM) mode in this waveguide was $72\pm8$~dB/cm ($232\pm13$~dB/cm). The increased loss after waveguide transfer indicates that scattering loss at the horizontal surfaces is greater in the membrane/diamond system than on the original GaP wafer.  An increase in scattering loss is to be expected due to the larger refractive index change at the GaP/diamond interface, surface roughness on the diamond, and any damage/defects caused by the 45 hour HF etch.  Defects due to the HF etch can be observed under the optical microscope with defect densities increasing with etch time and HF concentration.  Such defects could be eliminated if the GaP wafer was first bonded to the diamond and a substrate etching technique was used in place of the epitaxial lift-off technique.

Previous work has shown that small-mode-volume photonic bandgap cavities with cavity quality factors $Q$ up to 1700 are attainable in a GaP membrane undercut on the original GaP substrate~\cite{Rivoire08a}.  The losses we measured for GaP waveguides on diamond correspond to a maximum possible cavity quality factor $Q = 2\pi n_\mathrm{GaP}/\lambda\alpha$ ($\alpha$ is the $1/e$ decay length) of 20,000 (6000) for the TE (TM) mode.  This is already high enough that it should be possible to realize a large enhancement of the NV center-photon coupling efficiency in a GaP-based microcavity structure.  Finite element simulations (COMSOL)~\cite{ref:Kippenberg2004dem, resonatorsim} show that a GaP ring cavity with a $Q$ of 20,000 (limited by $\alpha$ and not radiation loss) is enough to significantly enhance the NV emission into the ZPL line.  The total spontaneous emission enhancement factor, in the weak-coupling limit and neglecting dispersion, is~\cite{SEfactor}
\small
\begin{equation}
F_{SE} = \frac{4 g^2}{\kappa \gamma_{\mathrm{total}}} =
\frac{3}{4 \pi^2}
\frac{\lambda^3}{n_{\mathrm{GaP}}^3}\frac{Q}{V}
\frac{n_{\mathrm{GaP}}}{n_\mathrm{D}}
\left|
\frac{\vec{E}(\vec{r}_{\mathrm{NV}})}{\vec{E}(\vec{r}_{\mathrm{max}})}
\right|^2
\frac{\gamma_{\mathrm{ZPL}}}{\gamma_{\mathrm{total}}}
\label{eq:F}
\end{equation}
\normalsize
where $g$ is the single-photon coupling strength, $\kappa$ is the cavity photon decay rate, $\gamma_{\mathrm{total}}=13\,\mathrm{MHz}$~\cite{ref:collins1983lum} is the total spontaneous emission rate including emission into the phonon sidebands, $V=18~(\lambda/n_\mathrm{GaP})^3$ is the traveling-wave mode volume obtained from the numerical simulation, $\vec{E}(\vec{r}_\mathrm{NV})$ is the electric field at the NV site, $ \vec{E}(\vec{r}_\mathrm{max})$ is the electric field maximum, and $\gamma_{\mathrm{ZPL}}=0.35 \, \mathrm{MHz}$ (derived from~\cite{ref:Davies1974vib}) is the spontaneous emission rate into the ZPL line. We find that an $F_\mathrm{SE}>1$ is possible for a ring diameter of 2.5~$\mathrm{\mu m}$ coupled to an NV center 20~nm beneath the diamond surface. This total $F_\mathrm{SE}$ factor corresponds to a 40-fold increase in emission into the zero phonon line.  Such an increase in ZPL emission would be valuable for repeat-until-success schemes based on the DLCZ protocol~\cite{ref:Duan2001lon} to create entanglement between two remote NV centers through interference and detection~\cite{Childress05a,Benjamin06a}.

\section{Supplemental material}
\subsection{Ring-cavity geometry}
%
%
\begin{figure}
{\includegraphics[height = 30 mm,
keepaspectratio]{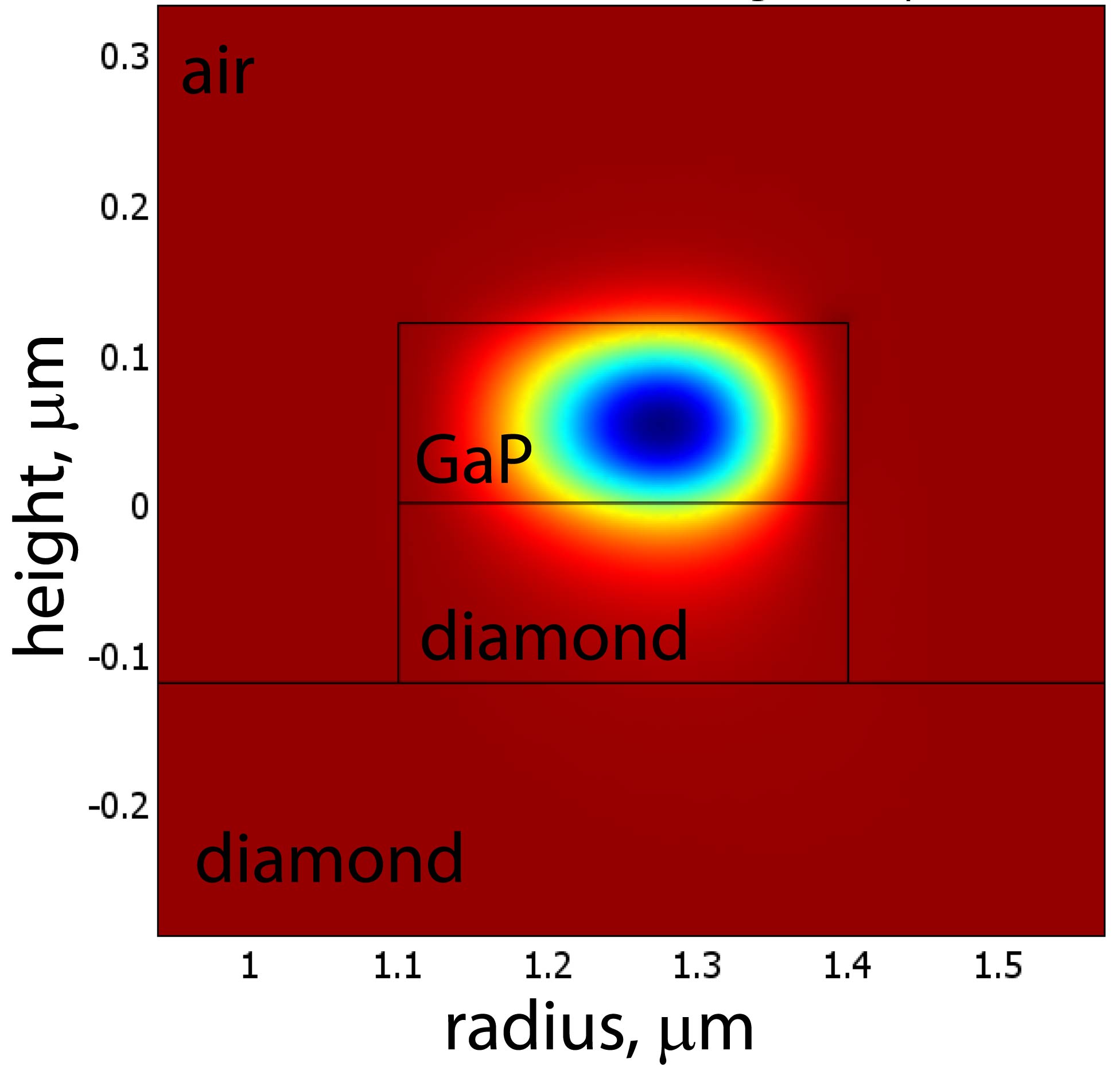}}\caption{Simulation of the electric field $|E|$ in a GaP/diamond ring cavity.} \label{fig:cavity}
\end{figure}
%
%
Fig.~\ref{fig:cavity} shows the geometry in the numerical simulation used to determine the radiation-limited $Q$ and mode volume of a GaP on diamond ring cavity.  The following parameters were used: GaP waveguide height = 120 nm, GaP waveguide width = 300~nm, GaP ring diameter = 2.5~$\mathrm{\mu m}$.  For better mode confinement the diamond is etched down 120 nm to form a ridge.  The NV is placed 20 nm beneath the GaP/diamond interface at the field maximum.

\subsection{Derivation of the spontaneous emission enhancement factor}

The expression given in Eq. 1 for the total on-resonance spontaneous emission enhancement factor in the weak-coupling limit is similar to the Purcell formula $F_\mathrm{SE} = 3 Q \lambda^3 / 4\pi^2V$ \cite{Purcell46a} except for some additional factors.

For a single atomic transition $i$, if we neglect dispersion in the dielectric, the ratio $g_i^2/\gamma_i$ given by Weisskopf-Wigner spontaneous emission theory is,
\begin{equation}
\frac{g_i^2}{\gamma_i} =
\frac{3}{16 \pi^2} \frac{\lambda^3}{n_{\mathrm{D}}^3}
\frac{\epsilon_{\mathrm{D}} |\vec{E}(\vec{r}_{\mathrm{NV}})|^2 \cos^2 \theta_i}
{\int d^3 \vec{r} \, \epsilon(\vec{r}) |\vec{E}(\vec{r})|^2}
\, \omega_i \, ,
\end{equation}
where $\vec{E}(\vec{r})$ is the electric field of the cavity mode of interest, $\epsilon(\vec{r})$ is the dielectric constant, $\vec{r}_{\mathrm{NV}}$ is the location of the NV center, $\omega_i$ is the frequency of the atomic transition, and $\theta_i$ is the angle between $\vec{E}(\vec{r}_{\mathrm{NV}})$ and the dipole moment of the atomic transition.  The factor $n_{\mathrm{D}}^3$ enters through the photon density of states in bulk diamond, the factor $\epsilon_{\mathrm{D}} = n_{\mathrm{D}}^2$ enters through the per-photon electric field strength in bulk diamond, and the integral in the denominator enters through the per-photon electric field strength of the cavity mode.  Combining this with a standard definition of mode volume,
\begin{equation}
V = \frac{\int d^3 \vec{r} \, \epsilon(\vec{r}) |E(\vec{r})|^2}
         { \epsilon(\vec{r}_\mathrm{max}) |E(\vec{r}_\mathrm{max})|^2 } \, ,
\end{equation}
and using $\epsilon(\vec{r}_{\mathrm{max}}) = n_{\mathrm{GaP}}^2$, $\gamma_i = \gamma_{\mathrm{ZPL}}$, $Q = \omega_i / \kappa$, and setting $\cos \theta_i = 1$ (optimal dipole alignment) we obtain the result in Eq. 1.  The factor $\gamma_{\mathrm{ZPL}}/\gamma_{\mathrm{total}}$ is important for NV centers since $<5\%$ of the total spontaneous emission is into the zero-phonon line.  If we omit this factor, we obtain instead the enhancement factor for the spontaneous emission rate through the zero-phonon line alone.

This work was supported by DARPA and the Air Force Office of Scientific Research through AFOSR Contract No.\ FA9550-07-C-0030.

\bibliography{kaimeibib4}

\begin{thebibliography}{21}
\expandafter\ifx\csname natexlab\endcsname\relax\def\natexlab#1{#1}\fi
\expandafter\ifx\csname bibnamefont\endcsname\relax
  \def\bibnamefont#1{#1}\fi
\expandafter\ifx\csname bibfnamefont\endcsname\relax
  \def\bibfnamefont#1{#1}\fi
\expandafter\ifx\csname citenamefont\endcsname\relax
  \def\citenamefont#1{#1}\fi
\expandafter\ifx\csname url\endcsname\relax
  \def\url#1{\texttt{#1}}\fi
\expandafter\ifx\csname urlprefix\endcsname\relax\def\urlprefix{URL }\fi
\providecommand{\bibinfo}[2]{#2}
\providecommand{\eprint}[2][]{\url{#2}}

\bibitem[{\citenamefont{Gaebel et~al.}(2006)\citenamefont{Gaebel, Domhan, Popa,
  Wittmann, Neumann, Jelezko, Rabeau, Stravrias, Greentree, Prawer
  et~al.}}]{Gaebel06a}
\bibinfo{author}{\bibfnamefont{T.}~\bibnamefont{Gaebel}},
  \bibinfo{author}{\bibfnamefont{M.}~\bibnamefont{Domhan}},
  \bibinfo{author}{\bibfnamefont{I.}~\bibnamefont{Popa}},
  \bibinfo{author}{\bibfnamefont{C.}~\bibnamefont{Wittmann}},
  \bibinfo{author}{\bibfnamefont{P.}~\bibnamefont{Neumann}},
  \bibinfo{author}{\bibfnamefont{F.}~\bibnamefont{Jelezko}},
  \bibinfo{author}{\bibfnamefont{J.~R.} \bibnamefont{Rabeau}},
  \bibinfo{author}{\bibfnamefont{N.}~\bibnamefont{Stravrias}},
  \bibinfo{author}{\bibfnamefont{A.~D.} \bibnamefont{Greentree}},
  \bibinfo{author}{\bibfnamefont{S.}~\bibnamefont{Prawer}},
  \bibnamefont{et~al.}, \bibinfo{journal}{Nature Phys.}
  \textbf{\bibinfo{volume}{2}}, \bibinfo{pages}{408} (\bibinfo{year}{2006}).

\bibitem[{\citenamefont{Jelezko et~al.}(2004)\citenamefont{Jelezko, Gaebel,
  Popa, Gruber, and Wrachtrup}}]{ref:jelezko2004obs}
\bibinfo{author}{\bibfnamefont{F.}~\bibnamefont{Jelezko}},
  \bibinfo{author}{\bibfnamefont{T.}~\bibnamefont{Gaebel}},
  \bibinfo{author}{\bibfnamefont{I.}~\bibnamefont{Popa}},
  \bibinfo{author}{\bibfnamefont{A.}~\bibnamefont{Gruber}}, \bibnamefont{and}
  \bibinfo{author}{\bibfnamefont{J.}~\bibnamefont{Wrachtrup}},
  \bibinfo{journal}{Phys. Rev. Lett.} \textbf{\bibinfo{volume}{92}},
  \bibinfo{pages}{076401} (\bibinfo{year}{2004}).

\bibitem[{\citenamefont{{Gurudev Dutt} et~al.}(2007)\citenamefont{{Gurudev
  Dutt}, Childress, Jiang, Togan, Maze, Jelezko, Zibrov, Hemmer, and
  Lukin}}]{GurudevDutt07a}
\bibinfo{author}{\bibfnamefont{M.~V.} \bibnamefont{{Gurudev Dutt}}},
  \bibinfo{author}{\bibfnamefont{L.}~\bibnamefont{Childress}},
  \bibinfo{author}{\bibfnamefont{L.}~\bibnamefont{Jiang}},
  \bibinfo{author}{\bibfnamefont{E.}~\bibnamefont{Togan}},
  \bibinfo{author}{\bibfnamefont{J.}~\bibnamefont{Maze}},
  \bibinfo{author}{\bibfnamefont{F.}~\bibnamefont{Jelezko}},
  \bibinfo{author}{\bibfnamefont{A.~S.} \bibnamefont{Zibrov}},
  \bibinfo{author}{\bibfnamefont{P.~R.} \bibnamefont{Hemmer}},
  \bibnamefont{and} \bibinfo{author}{\bibfnamefont{M.~D.} \bibnamefont{Lukin}},
  \bibinfo{journal}{Science} \textbf{\bibinfo{volume}{316}},
  \bibinfo{pages}{1312} (\bibinfo{year}{2007}).

\bibitem[{\citenamefont{Benjamin et~al.}(2006)\citenamefont{Benjamin, Browne,
  Fitzsimons, and Morton}}]{Benjamin06a}
\bibinfo{author}{\bibfnamefont{S.~C.} \bibnamefont{Benjamin}},
  \bibinfo{author}{\bibfnamefont{D.~E.} \bibnamefont{Browne}},
  \bibinfo{author}{\bibfnamefont{J.}~\bibnamefont{Fitzsimons}},
  \bibnamefont{and} \bibinfo{author}{\bibfnamefont{J.~J.~L.}
  \bibnamefont{Morton}}, \bibinfo{journal}{New J. of Phys.}
  \textbf{\bibinfo{volume}{8}}, \bibinfo{pages}{141} (\bibinfo{year}{2006}).

\bibitem[{\citenamefont{Childress et~al.}(2005)\citenamefont{Childress, Taylor,
  Sorensen, and Lukin}}]{Childress05a}
\bibinfo{author}{\bibfnamefont{L.}~\bibnamefont{Childress}},
  \bibinfo{author}{\bibfnamefont{J.~M.} \bibnamefont{Taylor}},
  \bibinfo{author}{\bibfnamefont{A.~S.} \bibnamefont{Sorensen}},
  \bibnamefont{and} \bibinfo{author}{\bibfnamefont{M.~D.} \bibnamefont{Lukin}},
  \bibinfo{journal}{Phys. Rev.~A} \textbf{\bibinfo{volume}{72}},
  \bibinfo{pages}{52330} (\bibinfo{year}{2005}).

\bibitem[{\citenamefont{Taylor et~al.}(2008)\citenamefont{Taylor, Cappellaro,
  Childress, Jiang, Budker, Hemmer, Yacoby, Walsworth, and
  Lukin}}]{ref:Taylor2008hig}
\bibinfo{author}{\bibfnamefont{J.~M.} \bibnamefont{Taylor}},
  \bibinfo{author}{\bibfnamefont{P.}~\bibnamefont{Cappellaro}},
  \bibinfo{author}{\bibfnamefont{L.}~\bibnamefont{Childress}},
  \bibinfo{author}{\bibfnamefont{L.}~\bibnamefont{Jiang}},
  \bibinfo{author}{\bibfnamefont{D.}~\bibnamefont{Budker}},
  \bibinfo{author}{\bibfnamefont{P.~R.} \bibnamefont{Hemmer}},
  \bibinfo{author}{\bibfnamefont{A.}~\bibnamefont{Yacoby}},
  \bibinfo{author}{\bibfnamefont{R.}~\bibnamefont{Walsworth}},
  \bibnamefont{and} \bibinfo{author}{\bibfnamefont{M.~D.} \bibnamefont{Lukin}},
  \bibinfo{journal}{Nature Phys.} \textbf{\bibinfo{volume}{Advance online
  publication}}, \bibinfo{pages}{doi:10.1038/nphys1075} (\bibinfo{year}{2008}).

\bibitem[{\citenamefont{Park et~al.}(2006)\citenamefont{Park, Cook, and
  Wang}}]{Park06a}
\bibinfo{author}{\bibfnamefont{Y.-S.} \bibnamefont{Park}},
  \bibinfo{author}{\bibfnamefont{A.~K.} \bibnamefont{Cook}}, \bibnamefont{and}
  \bibinfo{author}{\bibfnamefont{H.}~\bibnamefont{Wang}},
  \bibinfo{journal}{Nano Lett.} \textbf{\bibinfo{volume}{6}},
  \bibinfo{pages}{2075} (\bibinfo{year}{2006}).

\bibitem[{\citenamefont{Shen et~al.}(2008)\citenamefont{Shen, Sweeney, and
  Wang}}]{Shen08a}
\bibinfo{author}{\bibfnamefont{Y.}~\bibnamefont{Shen}},
  \bibinfo{author}{\bibfnamefont{T.~M.} \bibnamefont{Sweeney}},
  \bibnamefont{and} \bibinfo{author}{\bibfnamefont{H.}~\bibnamefont{Wang}},
  \bibinfo{journal}{Phys. Rev.~B} \textbf{\bibinfo{volume}{77}},
  \bibinfo{pages}{033201} (\bibinfo{year}{2008}).

\bibitem[{\citenamefont{Olivero et~al.}(2005)\citenamefont{Olivero, Rubanov,
  Reichart, Gibson, Huntington, Rabeau, Greentree, Salzman, Moore, Jamieson
  et~al.}}]{Olivero05a}
\bibinfo{author}{\bibfnamefont{P.}~\bibnamefont{Olivero}},
  \bibinfo{author}{\bibfnamefont{S.}~\bibnamefont{Rubanov}},
  \bibinfo{author}{\bibfnamefont{P.}~\bibnamefont{Reichart}},
  \bibinfo{author}{\bibfnamefont{B.~C.} \bibnamefont{Gibson}},
  \bibinfo{author}{\bibfnamefont{S.~T.} \bibnamefont{Huntington}},
  \bibinfo{author}{\bibfnamefont{J.}~\bibnamefont{Rabeau}},
  \bibinfo{author}{\bibfnamefont{A.~D.} \bibnamefont{Greentree}},
  \bibinfo{author}{\bibfnamefont{J.}~\bibnamefont{Salzman}},
  \bibinfo{author}{\bibfnamefont{D.}~\bibnamefont{Moore}},
  \bibinfo{author}{\bibfnamefont{D.~N.} \bibnamefont{Jamieson}},
  \bibnamefont{et~al.}, \bibinfo{journal}{Adv. Mater.}
  \textbf{\bibinfo{volume}{17}}, \bibinfo{pages}{2427} (\bibinfo{year}{2005}).

\bibitem[{\citenamefont{Wang et~al.}(2007)\citenamefont{Wang, Hanson,
  Awschalom, Hu, Feygelson, Yang, and Butler}}]{Wang07a}
\bibinfo{author}{\bibfnamefont{C.~F.} \bibnamefont{Wang}},
  \bibinfo{author}{\bibfnamefont{R.}~\bibnamefont{Hanson}},
  \bibinfo{author}{\bibfnamefont{D.~D.} \bibnamefont{Awschalom}},
  \bibinfo{author}{\bibfnamefont{E.~L.} \bibnamefont{Hu}},
  \bibinfo{author}{\bibfnamefont{T.}~\bibnamefont{Feygelson}},
  \bibinfo{author}{\bibfnamefont{J.}~\bibnamefont{Yang}}, \bibnamefont{and}
  \bibinfo{author}{\bibfnamefont{J.~E.} \bibnamefont{Butler}},
  \bibinfo{journal}{Appl. Phys. Lett.} \textbf{\bibinfo{volume}{91}},
  \bibinfo{pages}{201112} (\bibinfo{year}{2007}).

\bibitem[{\citenamefont{Yablonovitch et~al.}(1990)\citenamefont{Yablonovitch,
  Hwang, Gmitter, Florez, and Harbison}}]{Yablonovitch90a}
\bibinfo{author}{\bibfnamefont{E.}~\bibnamefont{Yablonovitch}},
  \bibinfo{author}{\bibfnamefont{D.}~\bibnamefont{Hwang}},
  \bibinfo{author}{\bibfnamefont{T.~J.} \bibnamefont{Gmitter}},
  \bibinfo{author}{\bibfnamefont{L.~T.} \bibnamefont{Florez}},
  \bibnamefont{and} \bibinfo{author}{\bibfnamefont{J.~P.}
  \bibnamefont{Harbison}}, \bibinfo{journal}{Appl. Phys. Lett.}
  \textbf{\bibinfo{volume}{56}}, \bibinfo{pages}{2419} (\bibinfo{year}{1990}).

\bibitem[{\citenamefont{Davies et~al.}(1992)\citenamefont{Davies, Lawson,
  Collins, Mainwood, and Sharp}}]{Davies92a}
\bibinfo{author}{\bibfnamefont{G.}~\bibnamefont{Davies}},
  \bibinfo{author}{\bibfnamefont{S.~C.} \bibnamefont{Lawson}},
  \bibinfo{author}{\bibfnamefont{A.~T.} \bibnamefont{Collins}},
  \bibinfo{author}{\bibfnamefont{A.}~\bibnamefont{Mainwood}}, \bibnamefont{and}
  \bibinfo{author}{\bibfnamefont{S.~J.} \bibnamefont{Sharp}},
  \bibinfo{journal}{Phys. Rev.~B} \textbf{\bibinfo{volume}{46}},
  \bibinfo{pages}{13157} (\bibinfo{year}{1992}).

\bibitem[{\citenamefont{Loudon}(1964)}]{ref:Loudon1964ram}
\bibinfo{author}{\bibfnamefont{R.}~\bibnamefont{Loudon}},
  \bibinfo{journal}{Advances in Physics} \textbf{\bibinfo{volume}{13}},
  \bibinfo{pages}{423} (\bibinfo{year}{1964}).

\bibitem[{\citenamefont{Rivoire et~al.}(2008)\citenamefont{Rivoire, Faraon, and
  Vuckovic}}]{Rivoire08a}
\bibinfo{author}{\bibfnamefont{K.}~\bibnamefont{Rivoire}},
  \bibinfo{author}{\bibfnamefont{A.}~\bibnamefont{Faraon}}, \bibnamefont{and}
  \bibinfo{author}{\bibfnamefont{J.}~\bibnamefont{Vuckovic}},
  \bibinfo{journal}{Appl. Phys. Lett.} \textbf{\bibinfo{volume}{93}}
  (\bibinfo{year}{2008}).

\bibitem[{\citenamefont{Kippenberg et~al.}(2004)\citenamefont{Kippenberg,
  Spillane, and Vahala}}]{ref:Kippenberg2004dem}
\bibinfo{author}{\bibfnamefont{T.}~\bibnamefont{Kippenberg}},
  \bibinfo{author}{\bibfnamefont{S.}~\bibnamefont{Spillane}}, \bibnamefont{and}
  \bibinfo{author}{\bibfnamefont{K.~J.} \bibnamefont{Vahala}},
  \bibinfo{journal}{Applied Physics Letters} \textbf{\bibinfo{volume}{85}},
  \bibinfo{pages}{6113} (\bibinfo{year}{2004}).

\bibitem[{res()}]{resonatorsim}
\bibinfo{note}{For specific geometry used in the simulation see supplemental
  materials (EPAPS)}.

\bibitem[{SEf()}]{SEfactor}
\bibinfo{note}{For full derivation see supplemental materials (EPAPS)}.

\bibitem[{\citenamefont{Collins et~al.}(1983)\citenamefont{Collins, Thomaz, and
  Jorge}}]{ref:collins1983lum}
\bibinfo{author}{\bibfnamefont{A.~T.} \bibnamefont{Collins}},
  \bibinfo{author}{\bibfnamefont{M.}~\bibnamefont{Thomaz}}, \bibnamefont{and}
  \bibinfo{author}{\bibfnamefont{M.}~\bibnamefont{Jorge}},
  \bibinfo{journal}{Phys. C, Solid State Phys.} \textbf{\bibinfo{volume}{16}},
  \bibinfo{pages}{2177} (\bibinfo{year}{1983}).

\bibitem[{\citenamefont{Davies}(1974)}]{ref:Davies1974vib}
\bibinfo{author}{\bibfnamefont{G.}~\bibnamefont{Davies}}, \bibinfo{journal}{J.
  Phys.C: Solid State Phys.} \textbf{\bibinfo{volume}{7}},
  \bibinfo{pages}{3797} (\bibinfo{year}{1974}).

\bibitem[{\citenamefont{Duan et~al.}(2001)\citenamefont{Duan, Lukin, Cirac, and
  Zoller}}]{ref:Duan2001lon}
\bibinfo{author}{\bibfnamefont{L.-M.} \bibnamefont{Duan}},
  \bibinfo{author}{\bibfnamefont{M.~D.} \bibnamefont{Lukin}},
  \bibinfo{author}{\bibfnamefont{J.~I.} \bibnamefont{Cirac}}, \bibnamefont{and}
  \bibinfo{author}{\bibfnamefont{P.}~\bibnamefont{Zoller}},
  \bibinfo{journal}{Nature} \textbf{\bibinfo{volume}{414}},
  \bibinfo{pages}{413} (\bibinfo{year}{2001}).

\bibitem[{\citenamefont{Purcell}(1946)}]{Purcell46a}
\bibinfo{author}{\bibfnamefont{E.}~\bibnamefont{Purcell}},
  \bibinfo{journal}{Phys. Rev.} \textbf{\bibinfo{volume}{69}},
  \bibinfo{pages}{681} (\bibinfo{year}{1946}).

\end{thebibliography}
\bibliographystyle{apsrev}

\end{document}